Louis Mello

# Fat-Tailed Distributions and Lévy Processes

The notion that natural disasters can be controlled is, of course, farcical; history is permeated with examples of countless failed attempts at this pointless task; it is synonymous with trying to build a perpetual motion machine. Nonetheless, there are ways to reduce their impact on human communities, particularly by looking away from the normal hypothesis.

8/16/2008

## *Introduction*

By definition natural catastrophes are extreme events, i.e., they are events that possess, at times, an infinitesimal likelihood of occurrence. In statistical terms this means that they usually occur on the tails of a probability distribution. This, in turn, elevates the importance of a correct measurement of the distribution's tails, not to mention the accurate identification of the distribution to which the data observed from nature belongs.

When we refer to extreme events we are, in fact, speaking of their placement in time, or, the fact that they occur extremely infrequently. As a result, the need to be able to predict such episodes with proper precision is of overwhelming importance to human communities that live under the constant threat of a major natural disaster such as the tsunamis and hurricanes that have shaken the world over the past few years. Obviously, we cannot control these events; however, we can better prepare ourselves for them through the discriminating use of fat-tailed statistical methods which come under broad scope of Lévy processes.

For many reasons that are widely known, the scientific community has come to rely over-abundantly on the Gaussian or *Normal* distribution. Ease of computation and a vast array of analyses are just the more self-evident reasons. The computer and the ability to generate rapid simulations have helped bridge the gap between the analytical and the experimental. Also noteworthy is the greater presence of more advanced studies of these complex processes, which result from the use of modern technology.

It is clear, regardless of how enamored one is with the Gaussian distribution and its attendant ease of use, limitations notwithstanding, that there is an urgent necessity to explore the notion that other processes may well be better suited for determining probability estimates for extreme events.

This monograph does not venture experimentation due to the author's limited experience with natural disasters. However, the mathematics show that there is a clear and proven way to obtain the parameters that would allow one to proceed with such experimentation.

It is in this spirit, as well as in the hope that other scientists will take on the task of performing detailed testing on natural phenomena, that this non-trivial recommendation is made.

## *Stochastic Lévy Processes – A Basic Definition*

Let us define a probability space as $(\Omega, \mathcal{F}, P)$ where $\Omega$ is the sample space of all conceivable results, $\mathcal{F}$ is a $\sigma-$algebra of the subsets of $\Omega$ and $P$ is the positive measure of total mass 1 on $(\Omega, \mathcal{F})$, which is essentially a probability. A stochastic process is then defined as a related group of random variables $(X(t), t \geq 0)$ defined on $(\Omega, \mathcal{F}, P)$ and taking values from a space $(E, \mathcal{E})$. We will, from now on, refer to *E* as a state space.

It is now easy to see that every $X(t)$ is represented by a corresponding metric mapping $(\mathcal{F}, \mathcal{E})$ from $\Omega$ to *E* and, hence, is really nothing more than an aleatory observation made on *E* at some *t*. We will also allow that *E* is a Euclidean space $\mathbb{R}^d$.

Let, for now, $E = \mathbb{R}^d$. A stochastic Lévy process $X = (X(t), t \geq 0)$ must satisfy:

1. *X* possesses independent and stationary increments,

2. $P[X(0) = 0] = 1$

3. *X* is a continuous process, hence, for all $a > 0$ and for all $s \geq 0$,
$\lim_{t \to s}(P \mid X(t) - X(s) \mid > a) = 0$.

The Lévy-Khintchine equation permits the generalization of a mapping $\phi_t : \mathbb{R}^d \to \mathbb{C}$ which yields:

$$\phi_t(t) = \mathbf{E}\left(e^{iu \cdot X(t)}\right) = \int_{\mathbb{R}^d} e^{iu \cdot y} p_t(dy) \qquad (1.1)$$

Where $p_t$ is the distribution of $X(t)$, i.e., $p_t = P \circ X(t)^{-1}$ and **E** is the mathematical expectation. $\phi_t$ must be continuous and positive definite. We remind the reader: Bochner's theorem states that any and all continuous positive definite mappings from $\mathbb{R}^d$ to $\mathbb{C}$ are Fourier transforms of finite measures on $\mathbb{R}^d$.

Then, given axiom 1, we can state that each $X(t)$ is infinitely divisible. For each $n \in \mathbb{N}$ there exists a probability measure $p_{t,n}$ on $\mathbb{R}^d$ with characteristic function $\phi_{t,n}$ such that $\phi_t(u) = (\phi_{t,n}(u))^n$ for each $u \in \mathbb{R}^d$.

This leads us to the fundamental theorem of infinitely divisible probability measures. The Lévy-Khintchine theorem can be written:

**Theorem 1.** *If $X = (X(t), t \geq 0)$ is a Lévy process, then $\phi_t(u) = e^{t\eta(u)}$ for each $t \geq 0$, $u \in \mathbb{R}^d$ where*

$$\eta(u) = ib \cdot u - \frac{1}{2} u \cdot au + \int_{\mathbb{R}^d - \{0\}} \left[ e^{iu \cdot y} - 1 - iu \cdot y 1_{\|y\|<1}(y) \right] \nu(dy) \qquad (1.2)$$

*for some $b \in \mathbb{R}^d$, a non-negative definite symmetric $d \times d$ matrix $\boldsymbol{a}$ and a Borel measure $\nu$ on $\mathbb{R}^d - \{0\}$ for which $\int_{\mathbb{R}^d - \{0\}} (\|y\|^2 \wedge 1) \nu(dy) < \infty$. Conversely, given a mapping if the form (1.2) it is always possible to construct a Lévy process for which $\phi_t(u) = e^{t\eta(u)}$.*

We call attention to the fact that the mapping $\eta : \mathbb{R}^d \to \mathbb{C}$ is, in fact, the characteristic exponent $\alpha$ of X. It is conditionally positive definite in that $\sum_{i,j=1}^{n} c_i \bar{c}_j \eta(u_i - u_j) \geq 0$, for all $n \in \mathbb{N}, c_1, \ldots, c_n \in \mathbb{C}$ with $\sum_{i=1}^{n} c_i = 0$.

## *Finding $\alpha$*

A strictly stable Lévy process is one for which a change in the time scale will have a similar effect on the spatial scale.

**Definition 1.** *Let $\mu$ be an infinitely divisible probability measure on $\mathbb{R}^d$. We will consider it **stable** if, for any $a > 0$ there are $b > 0$ and $c \in \mathbb{R}^d$ such that*

$$\hat{\mu}(z)^a = \hat{\mu}(bz) e^{i\langle c, z \rangle} \qquad (1.3)$$

*and we will consider it **strictly stable** if, for any $a > 0$ there is $b > 0$ such that*

$$\hat{\mu}(z)^a = \hat{\mu}(bz) \qquad (1.4)$$

*A **semi-stable** process is defined as some $a > 0$ where $a \neq 1$ and there are $b > 0$ and $c \in \mathbb{R}^d$ satisfying (1.3). We will call it **strictly semi-stable** if, for some $a > 0$ with $a \neq 1$, there is $b > 0$ satisfying (1.4).*

**Definition 2.** *Let $\{X_t : t \geq 0\}$ be a Lévy process on $\mathbb{R}^d$. It is called a stable, strictly stable, semi-stable or strictly semi-stable process if the distribution of $X_t$ at $t=1$ is respectively, stable, strictly stable, semi-stable or strictly semi-stable.*

**Example 1.** If $\mu$ is Gaussian on $\mathbb{R}^d$, then

$$\hat{\mu}(z) = e^{-\langle z, Az \rangle/2 + i\langle \gamma, z \rangle}$$

and $\mu$ is stable, as it satisfies (1.3) with $b = a^{\frac{1}{2}}$ and $c = \left(-a^{\frac{1}{2}} + a\right)\gamma$. Thus, if $\mu$ is Gaussian with a mean = 0, then it is strictly stable. If $\mu$ is Cauchy on $\mathbb{R}^d$ with parameter $c > 0$ and $\gamma \in \mathbb{R}^d$, then

$$\hat{\mu}(z) = e^{-c|z| + i\langle \gamma, z \rangle}$$

and $\mu$ is strictly stable, satisfying (1.4) with $b = a$.

If

$$\nu = \sum_{n=-\infty}^{\infty} b^{-n\alpha} \delta_{b^n x_0} \qquad (1.5)$$

with $b > 1$, $0 < \alpha < 2$, and $x_0 \in \mathbb{R}^d \setminus \{0\}$, then $\int (|x|^2 \wedge 1) \nu(dx) < \infty$ and the infinitely divisible distribution with generating triplet $(0, \nu, 0)$ is semi-stable with $a = b^\alpha$ in (1.3), but it is not stable. The semi-stability arises from

$$\hat{\mu}(z) = \exp\left[\sum_{-\infty}^{\infty} b^{-n\alpha} \left(e^{i\langle z, b^n x_0 \rangle} - 1 - i\langle z, b^n x_0 \rangle 1_D (b^n x_0)\right)\right]$$

**Definition 3.** *Let $\{X_t : t \geq 0\}$ be a stochastic process on $\mathbb{R}^d$. We shall call it **self-similar** if, for any $a > 0$, there is $b > 0$ such that*

$$\{X_{at} : t \geq 0\} \stackrel{d}{=} \{bX_t : t \geq 0\} \qquad (1.6)$$

*or it is called **latu sensu self-similar** if, for any $a > 0$, there are $b > 0$ and a function $c(t)$ from $[0, \infty)$ to $\mathbb{R}^d$ such that*

$$\{X_{at} : t \geq 0\} \stackrel{d}{=} \{bX_t + c(t) : t \geq 0\}. \qquad (1.7)$$

*It is called **semi self-similar** if, for some a > 0 with $a \neq 1$, there is b > 0 satisfying(1.6). It is called **latu sensu semi self-similar** if, for some a > 0 with $a \neq 1$, there are b > 0 and a function $c(t)$ satisfying(1.7).*

The link between these concepts and stable processes, etc. can be formed as follows:

**Proposition 1.** *Let $\{X_t : t \geq 0\}$ be a Lévy process on $\mathbb{R}^d$. Then it is self-similar, latu sensu self-similar, semi self-similar or latu sensu semi self-similar if and only if it is, respectively, strictly stable, stable, strictly semi-stable or semi-stable.*

**P R O O F**:

*Let $\mu = P_{X_1}$. Let us further assume that $\{X_t\}$ is latu sensu semi-stable. By definition there is a positive $a \neq 1$ for which (1.3) holds with some b and c. The Lévy processes $\{X_{at}\}$ and $\{bX_t + tc\}$ correspond to the distributions with characteristic functions $\hat{\mu}(z)^a$ and $\hat{\mu}(bz)e^{i\langle c,z \rangle}$, respectively. Hence, if $\{X_t\}$ and $\{X_t'\}$ are Lévy processes in law on $\mathbb{R}^d$ such that $P_{X_1} = P_{X_1'}$, then $\{X_t\}$ and $\{X_t'\}$ are identical in law and we can write*

$$\{X_{at}\} \stackrel{d}{=} \{bX_t + tc\} \tag{1.8}$$

*and hence $\{X_t\}$ is latu sensu semi self-similar. Conversely, if $\{X_t\}$ is latu sensu semi self-similar, then it follows from (1.7) that $P_{X_a} = P_{bX_1 + c(1)}$, i.e. $\hat{\mu}(z)^a = \hat{\mu}(bz)e^{i\langle c(1),z \rangle}$ and $\{X_t\}$ is semi stable. Simultaneously, we have shown that $c(t) = tc(1)$. The other assertions are proved similarly.*

Q.E.D.

**Theorem 2.** *Let $\{X_t : t \geq 0\}$ be a broad-sense self-similar or semi self-similar stochastically continuous, non-trivial process on $\mathbb{R}^d$ with $X_0 = $ const a.s. Denote by $\Gamma$ the set of all a > 0 such that there are b > 0 and $c(t)$ satisfying $\{X_{at} : t \geq 0\} \stackrel{d}{=} \{bX_t + c(t) : t \geq 0\}$. Then:*

i. *There is H > 0 such that, for every $a \in \Gamma$, $b = a^H$*

ii. *The set $\Gamma \cap (1, \infty)$ is non empty. Let $a_0$ be the infimum of this set.*

*If $a_0 > 1$, then $\Gamma = \{a_0^n : n \in \mathbb{Z}\}$ and $\{X_t\}$ is not broad-sense self-similar.*

*If $a_0 = 1$, then $\Gamma = (0, \infty)$ and $\{X_t\}$ is broad-sense self-similar.*

Consider a self-similar function $y(t)$. The difference between the maximum and minimum values of $y$ in a time interval $\Delta t$ defines a range for that interval, $R(\Delta t)$. Given that $y$ is self-similar, the ensemble-averaged value of $R$ will scale with $\Delta t$. We can write:

$$\langle R(\Delta t) \rangle = c \Delta t^H \tag{1.9}$$

where $c$ and $H$ are constants; $H$ defines the Hurst exponent[1]. For data that are only approximately self-similar, we use this relation to check their proximity to self-similarity, and also to obtain an effective value for $H$. We proceed as follows: create a moving window $\Delta t$ one point at a time through the raw data; an array of values $R(\Delta t)$ is created from which the mean $\langle R \rangle$ is found, thus reducing the effects of uneven sampling. This is repeated for a range of $\Delta t$ within the length of the data set. A plot of $\log \langle R(\Delta t) \rangle$ against $\log \Delta t$ will reveal any deviations from self-similarity, while the slope will yield the best estimate of $H$. Linear regression is utilized to calculate the 95% confidence interval for $H$.

Trivially, a function that is exactly constant over time has $H = 0$. At the other extreme, $H = 1$ indicates a function whose range increases linearly with time (for a positive $c$ in (1.9). Intermediate values of $H$ are generated by fractal functions: Random Gaussian noise possesses $H \approx 0.2$ while Gaussian Random Walks (whose next value in time is $y_{t-1} + \varepsilon$, where $\varepsilon$ is a random Gaussian increment) will yield $H \approx 0.5$. The value of $H$ does not uniquely establish correlation; however, uncorrelated series may present significant probabilities of observing greater values as the time-scale increases.

We now define $\alpha = \dfrac{1}{H}$ as a dimension in the probability space defined for the Generalized Lévy Characteristic Function (GLCF):

$$\log(f(t)) = i\sigma t - \gamma |t|^\alpha \left( 1 + i\beta \frac{t}{|t|} \tan\left( \alpha \frac{\pi}{2} \right) \right) \tag{1.10}$$

where:

$t$ = a constant of integration.

$\sigma$ = the location parameter of the mean.

$\gamma$ = is the scale parameter to adjust differences in time frequency of data.

$\beta$ = is the measure of skewness with $\beta$ ranging between -1 and +1.

$\alpha$ = the kurtosis and the fatness of the tails. Only when $\alpha = 2$ does the distribution become equal to the Gaussian distribution.

---

[1] Hurst, H. E. "*The Long Term Storage of Reservoirs*", *Transactions of the American Society of Engineers,* 116, 1951

## *Proof of the Relationship between $D_b$ and H*

The fractal dimension is related to the Hurst exponent by means of:

$$D_B = 2 - H \qquad (1.11)$$

**Proof:** Recall that to calculate the box counting dimension $D_B$ of a fractal object, we cover the object with $N$ boxes of side length $\delta$ and then compute $D_B$ using:

$$D_B = \frac{\log(N) - \log(V^*)}{\log\left(\frac{1}{\delta}\right)} \qquad (1.12)$$

Where $N$ = the number of boxes of length $\delta$ used to cover a line segment and $\frac{V^*}{\delta}$ is the minimum count of one dimensional boxes needed to cover said line segment.

For a fractional Brownian motion trace, suppose that we isolate a time series of $T$ time steps that spans 1 unit of time. During each time step of length $\frac{1}{T}$ the average vertical range of the function is $\left(\frac{1}{T}\right)^H$ due to the scaling properties of any self-similar process. In order to cover the plot of the function during a single time step, a rectangle of width $\frac{1}{T}$ and height $\left(\frac{1}{T}\right)^H$ is required.

The area of this rectangle is $\left(\frac{1}{T}\right)^{H+1}$, so the number of squares with side length $\frac{1}{T}$ needed to cover it is $\left(\frac{1}{T}\right)^{H-1}$. For all $T$ of the time steps, the total number of squares needed to cover the plot of the function is $\left(\frac{1}{T}\right)^{H-2}$. If we let $N = \left(\frac{1}{T}\right)^{H-2}$ and $\delta = \frac{1}{T}$, then the box counting dimension is given by:

$$D_B = \frac{\log\left(\frac{1}{T}\right)^{H-2}}{\log(T)} = 2 - H \qquad (1.13)$$

The time series spans 1 unit of time, so $V^* = 1^{D_B} = 1$.

This relationship between the fractal dimension and the Hurst exponent aligns perfectly with the notion of fractal dimension as a measure of the roughness of an object. As $H$ increases and the fractional Brownian motion displays greater persistence, the plot of the function becomes smoother and $D_B$ decreases accordingly. Conversely, as $H$ decreases and the fractional Brownian motion is more anti-persistent, the plot of the function becomes more jagged and $D_B$ increases.

The literature demonstrates the existence of a parametric function in which each coordinate's function is a Brownian motion trace. Similarly, we can construct fractional Brownian motion paths from traces. To calculate the box counting dimension of a fractional Brownian motion path with two coordinates, we examine a section of the path that results from $T$ time steps spanning 1 unit of time. During each time step of length $\frac{1}{T}$, each of the two traces has range $\left(\frac{1}{T}\right)^H$, so we can cover the path during a single time step with a square of side length $\left(\frac{1}{T}\right)^H$. For all $T$ time steps, the path requires $T$ such squares to cover it. Letting $N = T$ and $\delta = \left(\frac{1}{T}\right)^H$, we have:

$$D_B = \frac{\log(T)}{\log\left(\left(\frac{1}{T}\right)^H\right)} = \frac{1}{H} \quad (1.14)$$

Since $H$ can fall between 0 and 1, $\frac{1}{H}$ can assume values greater than 2. However, the fractal dimension of the path cannot exceed its Euclidean dimension $D_E$, so we modify the box counting dimension to be:

$$D_B = \min\left(\frac{1}{H}, D_E\right) \quad (1.15)$$

Thus, the fractal dimension of regular Brownian motion and anti-persistent fractional Brownian motion paths is 2, and the fractal dimension of persistent paths lies between 1 and 2.

## *Conclusion*

The above demonstrations provide us with the necessary mathematical support to state, unequivocally, that the fractal dimension of a time series allows us to estimate the characteristic exponent of a GLCF through its relationship with the Hurst exponent. One could, then, estimate either $H$ or $D_b$ and thus obtain $\alpha$.

**Comparison of Tail Probabilities between a Standard Normal and a Standardized Stable-Lévy Distribution (Table 1)**

| x | P[Normal] | P[Levy] | Ratio Levy/Normal |
|---|---|---|---|
| −10 | 7.619853E-24 | 9.861133E-05 | 12,941,368,742,522,800,000 |
| −9 | 1.128588E-19 | 0.000192105 | 1,702,174,781,955,010 |
| −8 | 6.220961E-16 | 0.000363951 | 585,039,308,533 |
| −7 | 1.280000E-12 | 0.000670638 | 523,935,763 |
| −6 | 9.865900E-10 | 0.001202084 | 1,218,423 |
| −5 | 2.866516E-07 | 0.00209627 | 7,313 |
| −4 | 3.167124E-05 | 0.003557187 | 112 |
| −3 | 0.001349898 | 0.006697816 | 4.96171990 |
| −2 | 0.022750132 | 0.020136505 | 0.88511595 |
| −1 | 0.158655254 | 0.117935844 | 0.74334660 |

*The table above provides ample proof of the massive magnitude of difference between the Gaussian and fat-tailed distributions. This means that the correct specification of the underlying distribution at the onset of the modeling effort is essential to the correct estimation of the intrinsic probabilities.*

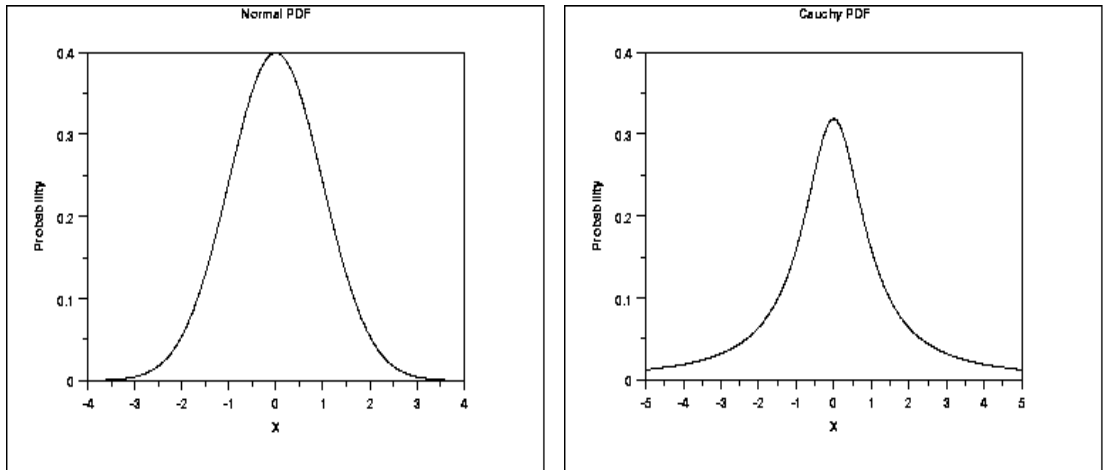

*An example of a curve with "fat tails" is the Cauchy distribution (a member of the stable-Pareto-Lévy class), shown above. In the case of the normal curve (on the left), the tails approach zero at -3.5 and 3.5 standard deviations. In the case of the Cauchy distribution the curve is still not even close to zero at -5 and 5 standard deviations. This illustrates the higher probability at the tail ends.*

The significance of fitting the appropriate distribution to the problem at hand cannot be over-emphasized. As can be seen from Table 1, an error in the estimate of the probability associated with an extreme event can be colossal. One of the major drawbacks with respect to fat-tailed distributions has always been the absence of a closed-form for parameterization. Also, depending on the value of $\alpha$ there may also be the issue of infinite variance.

There are several software implementations of recently defined algorithms based on the work of: Zolotarev (1986), Uchaikin and Zolotarev (1999), Christoph and Wolf (1992), Samorodnitsky and Taqqu (1994), Janicki and Weron (1994), and Nikias and Shao (1995) as well as the related topic of modeling with the extremes of data and heavy tailed distributions which is discussed in Embrechts et al. (1997), Adler et al. (1998), and in Reiss and Thomas (2001). These implementations allow the researcher to fit time series data to fat tailed stable Lévy class distributions yielding the parameters and the necessary modules for generating pseudo-random numbers from the observed distribution.

When the well-being and/or safety of whole populations lies in the balance, there should be no parsimony of effort in determining the best possible solution to a problem, as catastrophes like Katrina have so poignantly pointed out. There is no field of human endeavor that cannot benefit from studies predicated on the principles outlined in this paper, either directly or indirectly. It is in the interest of scientists and policymakers alike to pursue this course of research in order to better understand and integrate with the world which we all share.